\documentclass[sigconf]{acmart}

\acmConference[]{Preprint}{}{}
\acmBooktitle{Preprint}
\renewcommand\footnotetextcopyrightpermission[1]{}
\usepackage{booktabs}
\usepackage{multirow}
\usepackage{graphicx}
\usepackage{algorithm}
\usepackage{algpseudocode}
\usepackage{xcolor}
\usepackage{tikz}
\usepackage{caption}
\usepackage{tabularx}
\usepackage{makecell}
\usetikzlibrary{arrows.meta,positioning,shapes.geometric,fit}
\usepackage{enumitem}
\usepackage{array}
\newcolumntype{Y}{>{\raggedright\arraybackslash}X}


\begin{document}

\title{SlimPer: Make Personalization Model Slim and Smart}

\renewcommand{\authorsaddresses}{}

\author{Siqi Wang$^*$, Xianjie Chen$^{*\dagger}$, Shaofeng Deng, Albert Chen, Romil Shah, Jiawei Huang, Zhaoqin Wang, Zhang Zhang, Yiqun Liu, Meilei Jiang, Anish Dubey, Moyan Mei, Tongxin Wang, Nathan Berrebbi, Misael Manjarres, Armand Sauzay, Shardul Kothapalli, Aryaman Vinchhi, Kevin Johnstone, Juheon Lee, Gufan Yin, Ziheng Huang, Justin Lin, Mert Terzihan, Yilin Qi, Cynthia Yang, Colin Peppler, Qi Ding, Ruohan Sun, Ge Song, Litao Deng, Parichay Kapoor, Matt Ma, Huihui Cheng, Jiyuan Zhang, Yanli Zhao, Yiping Han, Fangqiu Han, Ning Yao, Arun Singh, Jordan Edwards, Zhengyu Su, Abhishek Kumar, Guangdeng Liao, Ankit Asthana}
\affiliation{\institution{Meta Platforms, Inc.}\country{USA}}
\renewcommand{\shortauthors}{Wang and Chen, et al.}
\begin{abstract}
Transformer-style architectures are increasingly adopted for industrial recommendation systems, yet they inherit a design premise misaligned with the task: generative models rely on per-token autoregressive prediction, which justifies maintaining large intermediate tensors that scale with sequence length. In contrast, recommendation systems produce a single set of relevance scores for each $\textit{<user, item>}$ pair without token-level supervision. Leveraging this observation, we propose \textbf{SlimPer}, which reformulates personalized ranking as \textit{iterative refinement of a compact, unified $\textit{<user, item>}$ knowledge base}. At each layer, the model selectively queries raw multi-modal user-side tokens, computes explicit relevance matching scores, and refines the knowledge base, all in $\mathcal{O}(N)$ per-layer cost with a fixed-size intermediate representation. As a result, model depth is decoupled from user history length, enabling deeper relevance understanding without proportional growth in compute or memory; request-only optimization further trims memory by sharing a single copy of user-side tokens across all candidate items. SlimPer unifies sparse, dense, and sequence features within a single backbone and provides inherent interpretability through its attention mechanism. Deployed on Instagram Reels and Feed, SlimPer yields measurable improvements in user engagement while streamlining the overall system and enabling effective modeling of 10k+ fine-grained user history events.
\end{abstract}

\maketitle
\makeatletter
\def\@oddhead{\normalfont\footnotesize\shorttitle\hfill\shortauthors}
\let\@evenhead\@oddhead
\def\@oddfoot{\normalfont\footnotesize\hfil\thepage\hfil}
\let\@evenfoot\@oddfoot
\makeatother
\renewcommand{\thefootnote}{}
\footnotetext{$^*$Equal contribution. $\dagger$Project lead and corresponding author (cxj@meta.com).}
\renewcommand{\thefootnote}{\arabic{footnote}}

\section{Introduction} \label{sec:intro}

\subsection{The Transformer Mismatch in Recommendation}

Discriminative personalization models, commonly known as ranking models, shape the content feeds, video recommendations, and product suggestions that users interact with every day. These models need to rank candidate items according to user preferences in real time while maintaining accurate predictions under production cost constraints.

Industrial recommendation models typically consume multi-modal input features, where each modality offers different trade-offs between computational cost and the ability to capture fine-grained information (see Table~\ref{tab:modalities} for a detailed comparison). In common practice, different modalities are processed by separate specialized components and then shallowly stitched together for final predictions. Recently, motivated by the success of Transformers in large language models (LLMs), adapting transformer-like architectures to recommendation systems has become a popular direction, either for modeling variable-length sequence features (e.g., HSTU~\cite{zhai2024hstu}, Interformer~\cite{zeng2024interformer}) or for unifying the modeling of different feature modalities (e.g., OneTrans~\cite{zhang2025onetrans}, HHFT~\cite{yu2025hhft}, RankMixer~\cite{zhu2025rankmixer}).

However, this trend overlooks a structural difference between recommendation systems and LLMs. In LLMs, per-position autoregressive supervision requires maintaining full token-level representations throughout the network, and the computational cost is amortized across predictions at every token position. \textit{Current recommendation systems have no such requirement}: they are trained for discriminative prediction and produce a single set of relevance scores for each $\textit{<user, item>}$ pair without token-level autoregressive losses. As a result, Transformer-like designs face an unnecessary dilemma: either intermediate representations are heavily compressed across layers, leading to information loss, or large intermediate tensors are preserved, incurring considerable memory and computational overhead as the number of input tokens scales. This dilemma is not inherent to the recommendation itself; it is an artifact of borrowing an architectural premise that does not apply.

\subsection{A New Paradigm: Iterative Refinement of a Knowledge Base}

We argue that the appropriate abstraction for discriminative recommendation is not sequence modeling, but \textit{iterative refinement} over a compact user-item knowledge base. Under this formulation, the model maintains a fixed-size representation of $\textit{<user, item>}$ relevance and progressively refines it by selectively querying raw multi-modal evidence, without materializing large intermediate tensors whose size grows with input sequence length.

\begin{figure}[t]
  \centering
  \includegraphics[width=\linewidth]{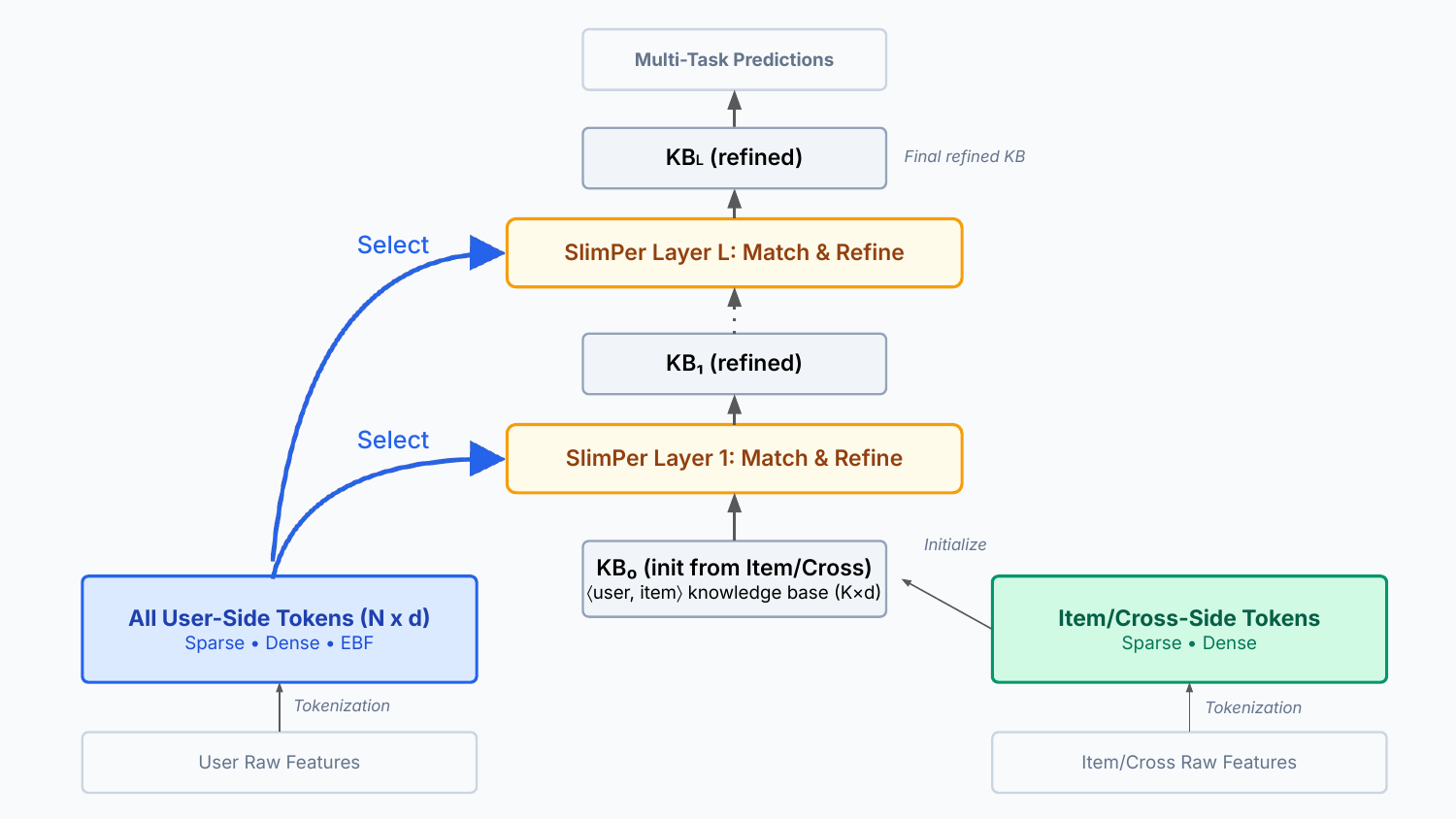}
  \caption{\textbf{High-level overview of SlimPer.} Three design principles: \textbf{(1) Slim} --- fixed-size knowledge base ($K \times d$), cost decoupled from input length $N$; \textbf{(2) Complete Access} --- user-side tokens queried at every layer; \textbf{(3) ROO-Aware} --- user-side tokens computed once and shared across all candidates in a request.}
  \label{fig:intro}
\end{figure}

Based on this principle, we propose \textbf{SlimPer}, a recommendation-centric architecture for personalized ranking. As shown in Fig.~\ref{fig:intro}, SlimPer formulates personalized ranking as an iterative refinement process over a unified fixed-size $\textit{<user, item>}$ knowledge base. Each layer performs three steps: (1) \textit{Selection}: all-modality-aware QKV attention queries the complete set of raw user-side tokens conditioned on the current knowledge base; (2) \textit{Matching}: explicit dot-product relevance scores are computed between the retrieved evidence and multifaceted templates; (3) \textit{Refinement}: the knowledge base is updated using the matched evidence.

This design resolves the transformer dilemma at its root: increasing model depth to improve understanding of the $\textit{<user, item>}$ pair no longer requires maintaining large intermediate tensors whose size scales with the number of user-side input tokens. At the same time, model quality is preserved because every layer retains direct access to the complete set of raw user-side information when refining the knowledge base. The resulting smaller intermediate representations translate directly into lower computational and memory cost (i.e., \textit{Slim}). Moreover, the knowledge base focuses its entire learning capacity on $\textit{<user, item>}$ relevance, allocating resources more deliberately toward relevance matching compared to architectures that must maintain full token-level representations (i.e., \textit{Smart}).

In addition, inspired by the success of request-only optimization (ROO) in HSTU~\cite{zhai2024hstu} for processing user-side sequence features, we explicitly separate user-side features from non-user-side features and extend ROO to all user-side feature modalities. This is another recommendation-centric design choice, motivated by the fact that recommendation models typically rank multiple candidate items for the same user during both training and inference. 

We conduct experiments on Instagram's two major ranking surfaces, Reels and Feed, using final-stage ranking models. SlimPer broadly outperforms prior models while achieving lower computational and memory cost with a more streamlined model architecture.

\subsection{Contributions}
In summary, our main contributions are as follows:

\begin{enumerate}[leftmargin=*] 

\item \textbf{Iterative refinement as a new paradigm for personalized ranking:} We formulate discriminative recommendation as an iterative refinement process over a compact $\textit{<user, item>}$ knowledge base, departing from the dominant transformer-mimicking paradigm. At each refinement step, the model directly queries the full set of raw multi-modal tokens, thereby decoupling model depth from input sequence length. This design achieves $\mathcal{O}(N)$ per-layer complexity with a fixed-size intermediate representation, effectively resolving the efficiency-quality trade-off faced by transformer-like architectures when scaling to long user histories.

\item \textbf{Unified multi-modal architecture with recommendation-centric efficiency:} SlimPer naturally provides a unified backbone for processing sparse, dense, and sequence features using only standard neural network building blocks including MLPs, QKV attention, linear projections, and RMSNorm~\cite{zhang2019rmsnorm}. Furthermore, inspired by the effectiveness of request-only optimization (ROO) in HSTU for user history modeling, SlimPer extends ROO awareness to all user-side feature modalities: user-side tokens are computed once per request batch and shared across all candidate items within the same request, reducing redundant computation and memory overhead. Together, these design choices improve both ranking quality and system efficiency compared with transformer-mimicking baselines.

\item \textbf{Intrinsic interpretability for real-world recommender systems:} The iterative refinement framework also provides intrinsic interpretability. At each layer, the QKV attention mechanism directly associates recommendation outcomes with specific user interaction history events, making model behavior easier to interpret and debug in real-world recommendation scenarios.

\end{enumerate}

\section{Methodology}
\label{sec:methodology}

\subsection{Design Philosophy}

The design principles motivating SlimPer arise from the observation that recommendation requires only discriminative prediction, not per-token autoregressive output. The idea of iterative refinement over a compact knowledge base, and the resulting architectural advantages (i.e., decoupled depth and length, complete information access, and more focused capacity allocation), were introduced in Section \ref{sec:intro}. We now formalize these principles into a concrete architecture.

\subsection{Problem Formulation}
Consider a recommendation request consisting of a user $u$ and a set of candidate items $\mathcal{I} = \{i_1, i_2, \ldots, i_C\}$ to be ranked. The input consists of diverse sets of feature modalities: a set of Sparse features ($\mathbf{S}$), Dense features ($\mathbf{D}$), and a sequence of user interaction history events ($\mathbf{E}$). The goal is to learn a scoring function $f(u, i; \mathbf{S}, \mathbf{D}, \mathbf{E}) \rightarrow \mathbb{R}^{\tau}$ that estimates the relevance of $i$ by matching it against the user's latent interests representation. In multi-task settings, this function simultaneously predicts $\tau$ objectives representing different engagement signals. 

\begin{table}[t]
\centering
\footnotesize
\caption{Different information modalities}
\label{tab:modalities}
\setlength{\tabcolsep}{6pt}
\renewcommand{\arraystretch}{1.1}
\begin{tabular}{@{}p{0.05\textwidth} p{0.1\textwidth} p{0.23\textwidth}}
\toprule
\textbf{Modality} & \textbf{Input Type} & \textbf{Representative Signals}\\
\midrule
Sparse &
High-cardinality categorical IDs &
Summarized user interests (e.g., top engaged creators or categories), providing salient but coarse-grained preference signals. \\

\addlinespace
Sequence &
Structured user engagement records &
Fine-grained behavioral signals capturing individual user actions (e.g., content consumed, dwell time, engagement type, and timestamp), providing high-resolution temporal information. \\

\addlinespace
Dense &
Continuous numerical values &
Contextual and real-time signals, such as user-level statistics (e.g., CTR), item quality metrics (e.g., reshare rate), and contextual attributes (e.g., time or device). \\

\bottomrule
\end{tabular}
\end{table}

\subsection{Notation and Preliminaries}

We restrict the model to a minimal set of building blocks: $\mathrm{MLP}$ (multilayer perceptron with SiLU activations), $\mathcal{L}$ (segment-wise linear projection: $\beta = \mathcal{L}(\alpha) = \rho \alpha$, $\rho \in \mathbb{R}^{L_{\mathrm{out}} \times L_{\mathrm{in}}}$), and the all-modality $\textit{<user, item>}$ knowledge base $\mathcal{X}$, initialized as $\mathcal{X}^{0} = \mathcal{L}(\mathbf{S}_{\text{in}})$ from non-user-side input tokens. The knowledge base has shape $\mathcal{X}^{k} \in \mathbb{R}^{K \times d}$, where $K$ is the number of knowledge base slots and $d$ is the embedding dimension. The size may in general differ across layers; in our experiments we keep it consistent with $K=64$ across all layers. Full notation details are in Appendix~\ref{app:notation}.

\subsection{Overall Architecture}

\begin{figure*}[t]
    \centering
    \includegraphics[width=0.85\textwidth]{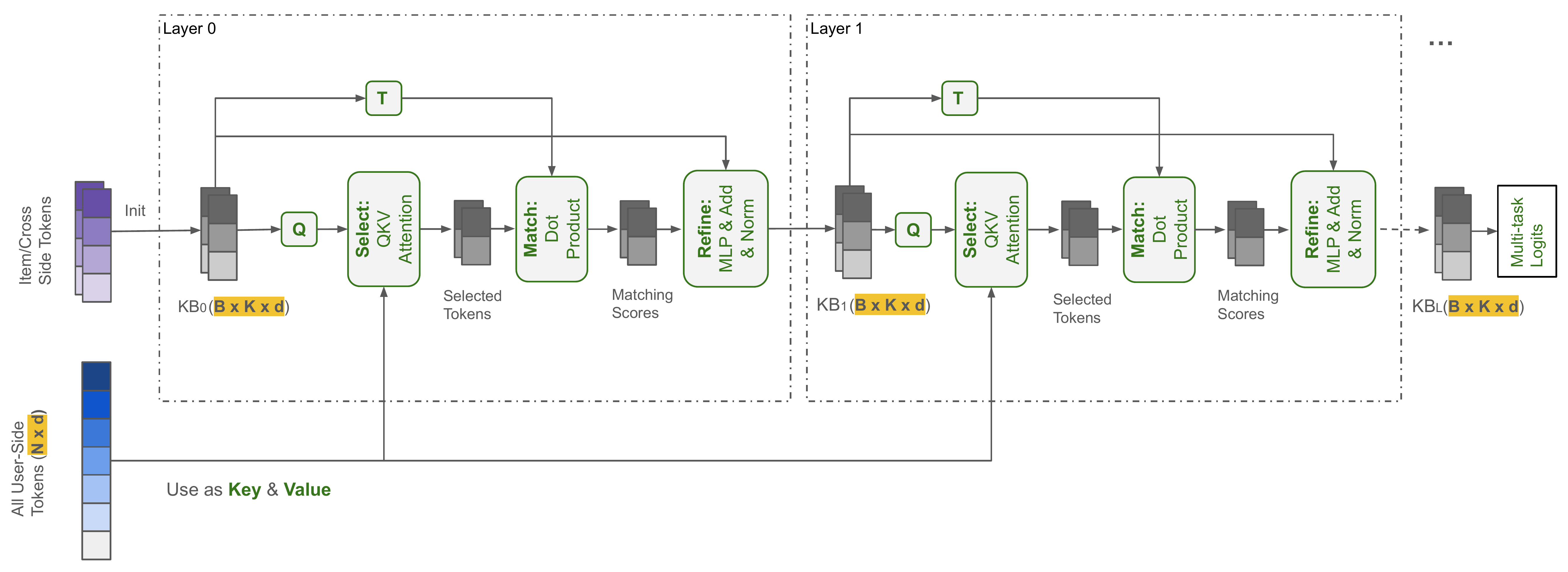}
    \caption{Architecture of the multi-layer SlimPer model. Each layer performs iterative refinement of a compact knowledge base ($K$=16--64 slots) by querying the full set of user-side tokens ($N$=$\mathcal{O}(10^3)$--$\mathcal{O}(10^4)$ in production), decoupling the memory and compute cost of stacking layers from input sequence length.}
    \label{fig:slimper_overall}
\end{figure*}

A SlimPer model consists of stacked SlimPer layers preceded by multi-modality tokenization. The overall architecture is shown in Fig.~\ref{fig:slimper_overall}. The model takes raw sparse and dense features, along with a sequence of user history events, as inputs. All features are first tokenized using their corresponding tokenizers (see Table~\ref{tab:continuous_tokenization}), producing a set of multi-head $d$-dimensional embeddings. For simplicity, we describe our algorithms in single-head form without loss of generality. Each layer performs the Select--Match--Refine cycle (Section~\ref{sec:slimper_layer}), leveraging standard Query-Key-Value (QKV) attention for all-modality-aware token selection and explicit relevance matching.

The model inherently incorporates request-only optimization (ROO) awareness by explicitly decoupling user-side and item-side features. Historically, ROO has been associated with quality-efficiency trade-offs (e.g., two-tower architectures in early-stage ranking) or narrowly applied to user history modeling. Our design extends ROO awareness across the entire model, encompassing all user-side input features beyond just user history. Tokenization for all user-side features is performed once per request batch, and a single copy of these user-side tokens is shared across all items within the same request, minimizing memory and compute cost.

\begin{figure}[H]
    \centering
    \includegraphics[width=0.85\linewidth]{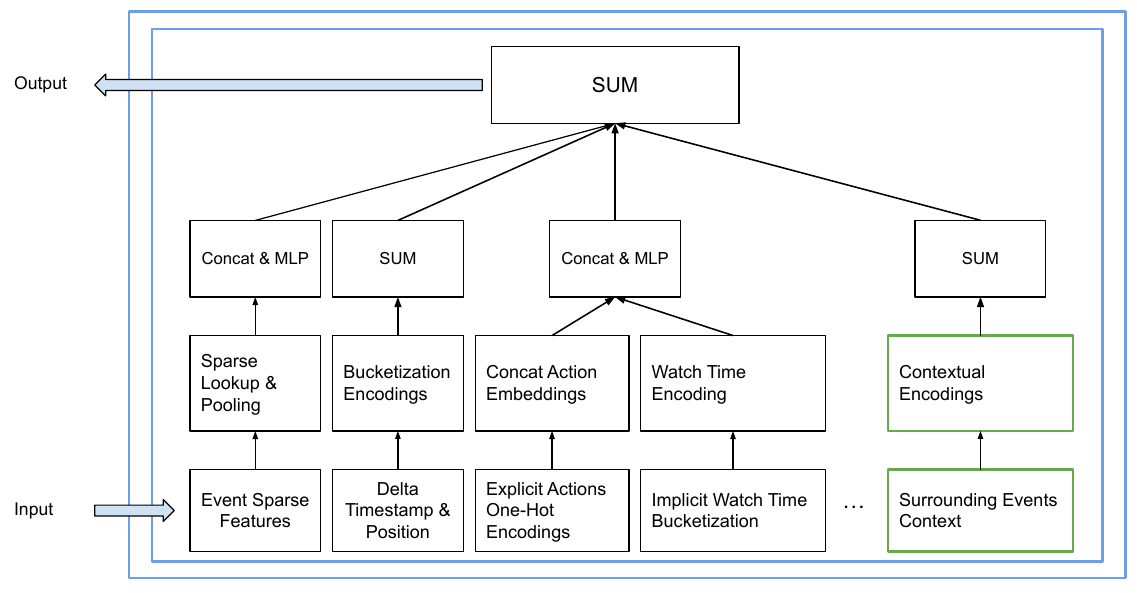}
    \caption{Event modeling module that encodes each event using content, engagement type, temporal and contextual information.}
    \label{fig:em_module_context}
\end{figure}

\subsection{Tokenization Module}
\label{sec:tokenization}

All raw input features are converted into modality-specific tokens within the unified backbone. Sparse features are tokenized via embedding pooling into fixed-size vectors. Each user history event is encoded into a $1 \times d$ token by fusing content, engagement type, temporal, and contextual information. Dense features are concatenated and projected through an MLP. The resulting multi-modal token set serves as input to the SlimPer layers. Full tokenization details for standard components (embedding pooling and dense processing) are provided in Appendix~\ref{app:tokenization}.

\paragraph{Contextual Encoding for Event Tokens.}
Intuitively, user behavior often exhibits local continuity, where an event is most informative when interpreted together with its nearby events. To capture these local contextual relationships, we introduce an additional contextual encoding $\mathbf{e}_{\text{context}}^{(t)}$ that explicitly summarizes neighborhood information around each event token (Fig.~\ref{fig:em_module_context}, green box):

\begin{equation}
 \mathbf{e}_{\text{context}}^{(t)} = \Psi_{\text{context}}(x_{t-w+1}, \ldots, x_{t})
\end{equation}

where $\Psi_{\text{context}}$ represents the encoding function, $w$ is the window size, and $x_{t}$ denotes the information collected from the user history event at position $t$. We explored convolutional neural networks~\cite{lecun1998gradientbased}, sliding-window attention~\cite{beltagy2020longformer}, and combinations of both as $\Psi_{\text{context}}$; all variants achieved similar performance. We also experimented with using different information at each event position for contextual encoding; see the ablation study in Section~\ref{sec:contextual_effect} for details.

\subsection{SlimPer Layer: Select, Match, and Refine}
\label{sec:slimper_layer}

As illustrated in Fig.~\ref{fig:slimper_overall}, each SlimPer layer follows a three-step iterative refinement procedure. Tokens from all modalities are jointly processed in Steps 1 and 2, where QKV attention is used to retrieve relevant user-side tokens and compute explicit relevance matching scores. Step 3 then refines the knowledge base with the matched evidence. This design reallocates compute and memory away from pairwise user-side token-to-token interactions and toward learning discriminative user-item interactions. Table~\ref{tab:complexity} compares the compute and memory allocation of SlimPer with representative transformer-like architectures.

\begin{table*}[t]
\centering
\scriptsize
\setlength{\tabcolsep}{3pt}
\renewcommand{\arraystretch}{1.1}
\caption{Per-item Big-O complexity across architectures. The rightmost column is per-layer user-item interaction capacity, where SlimPer's $K\times$ advantage is analyzed in Section~\ref{sec:stacking}. Typical production values: $N$=$\mathcal{O}(10^3)$--$\mathcal{O}(10^4)$, $K$=64, $q$=16, $B \approx 5$ (training) / $\mathcal{O}(10^2)$ (inference).}\label{tab:complexity}
\begin{tabularx}{\textwidth}{@{}lXXX@{}}
\toprule
\textbf{Model} &
\textbf{Tokenization Cost (per item)} &
\textbf{Stacked Layers Cost (per item)} &
\textbf{User-Item Interaction Capacity} \\
\midrule
\makecell[l]{\textbf{Conventional DLRM Model}\\$\bullet$ \emph{not ROO-aware; no sequence feature support}} &
Memory: $\mathcal{O}(N)$ \newline
Compute: $\mathcal{O}(N)$ &
Memory: $\mathcal{O}(L \cdot 64)$ \newline
Compute: $\mathcal{O}(L \cdot N \cdot 16)$ &
$\mathcal{O}(L \cdot N \cdot 64)$ \\
\addlinespace[2pt]
\makecell[l]{\textbf{ROO-aware Transformer-like Model (e.g., HSTU)}\\$\bullet$ \emph{ROO-aware for user history modeling}} &
Memory: $\mathcal{O}(N/B)$ \newline
Compute: $\mathcal{O}(N/B)$ &
Memory: $\mathcal{O}(L \cdot N / B)$ \newline
Compute: $\mathcal{O}\big(L \cdot (N+1)^2 / B\big)$ &
$\mathcal{O}(L \cdot N)$ \\
\addlinespace[2pt]
\makecell[l]{\textbf{Non-ROO-aware Transformer-like Model (e.g., OneTrans)}\\$\bullet$ \emph{not ROO-aware}} &
Memory: $\mathcal{O}(N)$ \newline
Compute: $\mathcal{O}(N)$ &
Memory: $\mathcal{O}(L \cdot N)$ \newline
Compute: $\mathcal{O}\big(L \cdot N^2\big)$ &
$\mathcal{O}(L \cdot N)$ \\
\addlinespace[2pt]
\makecell[l]{\textbf{Sub-quadratic Transformer-like Model}\\(e.g., linear or sliding-window attention)\\$\bullet$ \emph{linearized attention}} &
Memory: $\mathcal{O}(N)$ \newline
Compute: $\mathcal{O}(N)$ &
Memory: $\mathcal{O}(L \cdot N)$ \newline
Compute: $\mathcal{O}(L \cdot N)$ &
$\mathcal{O}(L \cdot N)$ \\
\addlinespace[2pt]
\makecell[l]{\textbf{SlimPer}\\$\bullet$ \emph{ROO-aware for all feature modalities}} &
Memory: $\mathcal{O}(N/B)$ \newline
Compute: $\mathcal{O}(N/B)$ &
Memory: $\mathcal{O}(L \cdot K)$ \newline
Compute: $\mathcal{O}(L \cdot N \cdot q)$ &
$\mathcal{O}(L \cdot N \cdot K)$ \\
\bottomrule
\end{tabularx}
\end{table*}

\subsubsection{Step 1: Token Selection via All-Modality-Aware Attention}

We leverage information from all input feature modalities during attention-based token selection. Let $\mathcal{X}^{k}$ denote the unified knowledge base at iteration $k$, representing the model's current understanding of the user-item pair. Query vectors $\mathbf{Q}$ are derived from $\mathcal{X}^{k}$ through segment-wise linear projection:

\begin{align}
    \mathbf{Q} &= \mathcal{L}(\mathcal{X}^{k}) \in \mathbb{R}^{q \times d}.
\end{align}

Intuitively, $\mathbf{Q}$ captures the information that the model seeks to retrieve from user-side tokens, conditioned on the current user-item context encoded in $\mathcal{X}^{k}$. All user-side input tokens serve as keys $\mathbf{K}$ and values $\mathbf{V}$ (with optional linear projection). We organize them into two key-value sets: fixed-size sparse features and variable-length sequence features. QKV attention is applied to both sets to compute attention weights, which are used to aggregate the corresponding values to produce attention output $\mathbf{R}$:

\begin{equation}
    \mathbf{R} = \Phi(\mathbf{Q}, \mathbf{K}) \, \mathbf{V}
\end{equation}
Here the attention kernel $\Phi$ is modality-dependent. For fixed-size sparse tokens, we employ a lightweight parametric kernel implemented as an $\mathrm{MLP}$ since the token count is constant. For variable-length user history sequence features, we adopt a non-parametric standard scaled dot-product kernel~\cite{vaswani2017attention}. Dense features, which typically have relatively small dimensionality (e.g., $\sim$1000 floats), are directly propagated and fused at each layer without passing through an attention kernel:

\begin{equation}
    \Phi(\mathbf{Q}, \mathbf{K}) =
    \begin{cases}
    \mathrm{MLP}(\operatorname{Concat}(\mathcal{L}(\mathbf{Q}), \mathcal{L}(\mathbf{K}))), & \text{sparse features} \\
    \text{Softmax}\left(\gamma QK^\top\right), & \text{sequence features}
    \end{cases}
\end{equation}

where $\gamma$ denotes the softmax temperature. Applying this formulation to the two user-side token streams yields two attention-based representations: 
\begin{align}
    \mathbf{R}_{s} &= \Phi_{s}(\mathbf{Q}, \mathbf{S}) \, \mathbf{S}, & \mathbf{R}_{s} &\in \mathbb{R}^{q \times d}, \\
    \mathbf{R}_{e} &= \Phi_{e}(\mathbf{Q}, \mathbf{E}) \, \mathbf{E}, & \mathbf{R}_{e} &\in \mathbb{R}^{q \times d}.
\end{align}

The attention outputs $\mathbf{R}_{s}$ and $\mathbf{R}_{e}$ summarize the evidence selected from both token streams. In our implementation, the raw sparse feature tokens $\mathbf{S}$ and sequence feature tokens $\mathbf{E}$ are directly used as keys and values in the QKV attention. We also explored applying additional linear projections to derive keys and values, but observed no performance improvement; therefore, these projections are omitted for computational efficiency.

This design yields an all-modality-aware attention mechanism. Since the query representation $\mathbf{Q}$ is derived from the unified knowledge base $\mathcal{X}^{k}$, it incorporates information from all input modalities. Consequently, sparse feature information can influence token selection over sequence features, and vice versa.

\subsubsection{Step 2: Explicit Multifaceted Dot-Product Relevance Matching}

To perform relevance matching, we derive a set of multifaceted templates $\mathbf{T} \in \mathbb{R}^{t \times d}$ from the unified $\textit{<user, item>}$ knowledge base using a linear transformation:

\begin{equation}
    \mathbf{T} = \mathcal{L}(\mathcal{X}^{k}), \quad \mathbf{T} \in \mathbb{R}^{t \times d}.
\end{equation}

The relevance matching scores are computed via explicit dot products between the templates and each stream's attention output:

\begin{align}
    \boldsymbol{\lambda}_{s} &= \mathrm{DotProduct}(\mathbf{R}_{s}, \mathbf{T}), & \boldsymbol{\lambda}_{s} &\in \mathbb{R}^{q \times t}, \\
    \boldsymbol{\lambda}_{e} &= \mathrm{DotProduct}(\mathbf{R}_{e}, \mathbf{T}), & \boldsymbol{\lambda}_{e} &\in \mathbb{R}^{q \times t}.
\end{align}

\subsubsection{Step 3: Refine the Knowledge Base}

The relevance scores are then normalized and transformed via $\mathrm{MLP}_{\mu}$ to refine the $\textit{<user, item>}$ knowledge base $\mathcal{X}^{k}$ and produce the next layer's representation $\mathcal{X}^{k+1}$. The dense tokens $\mathbf{D}$ are concatenated to incorporate dense feature information when updating the knowledge base:

\begin{equation}
\begin{aligned}
    \mathcal{X}^{k+1} = \mathcal{X}^{k} + \mathrm{MLP}_{\mu}\Big(&\mathrm{Concat}\big(\mathrm{RMSNorm}(\boldsymbol{\lambda}_{s}), \, \mathrm{RMSNorm}(\boldsymbol{\lambda}_{e}), \\
    &\mathcal{L}(\mathcal{X}^{k}), \, \mathbf{D}\big)\Big)
\end{aligned}
\end{equation}

At each layer, we optionally output a task embedding for each prediction task, representing the model's current understanding of that task. The task embeddings across layers can be summed before applying a final linear projection to produce task logits:

\begin{equation}
    \mathbf{P}_{p}^{k} = \mathrm{MLP}_{p}(\mathcal{L}(\mathcal{X}^{k}))
\end{equation}

where $p \in [1, \tau]$ indexes the prediction tasks.

\subsection{Iterative Personalization via Stacking SlimPer Layers}
\label{sec:stacking}

SlimPer is inherently stackable: multiple SlimPer layers are composed sequentially, each performing one iteration of the Select--Match--Refine cycle (Section~\ref{sec:slimper_layer}) over the shared knowledge base $\mathcal{X}$. In our production setting, we typically stack 5 to 10 layers.

The intuition is that \textbf{deeper layers benefit from the evidence accumulated by earlier layers}, enabling progressively more informed personalization. For example, earlier layers may focus on collecting broad evidence related to the target item, such as whether the user historically engaged with dog videos. This accumulated evidence then conditions subsequent layers, which can focus on more nuanced signals, e.g., whether recent behaviors (e.g., frequent skipping of dog videos) indicate that the user's interest has shifted. In this way, later layers can determine whether earlier preferences should be reinforced or discounted.

This iterative refinement capability is enabled by two architectural properties. First, each layer directly attends to the \textit{complete set of raw input tokens} across all modalities, preserving fine-grained signals without incurring the memory overhead of maintaining large intermediate tensors. Unlike transformer-like architectures, which progressively compress information into intermediate hidden states, SlimPer performs lossless cross-modality and cross-side (user$\leftrightarrow$item) fusion at every depth. Second, the knowledge base maintains a fixed size throughout the refinement process. As a result, increasing depth does not require intermediate memory to scale with input length, making it practical to stack many refinement layers even for users with long interaction histories.

\subsubsection{Information-Theoretic Justification}The fixed-size knowledge base that enables these efficiency gains may appear to sacrifice information, but analysis reveals the opposite: SlimPer's bottleneck \textit{redirects} capacity toward task-relevant signals rather than discarding them.

\textit{The bottleneck is capacity-abundant.}
The recommendation task demands only $\tau$ scalar predictions per $\textit{<user, item>}$ pair, so the mutual information between target $Y$ and input $X$ is bounded by $H(Y) \leq \tau$ bits---far below the representational capacity of a $K \times d$ knowledge base. Moreover, as shown in the interaction capacity column of Table~\ref{tab:complexity}, SlimPer achieves $\mathcal{O}(L \cdot N \cdot K)$ user-item interaction capacity---matching DLRM and $K\times$ larger than transformer-like models that achieve only $\mathcal{O}(L \cdot N)$. The difference arises because transformers allocate the majority of their $\mathcal{O}(N^2)$ computation to pairwise user-side token interactions that do not directly serve relevance estimation, whereas SlimPer directs all $K \times N$ interactions per layer toward knowledge-base-to-token relevance matching. The bottleneck thus constrains \textit{intermediate memory}, not \textit{learning capacity for user-item relevance}. Notably, this capacity argument is independent of whether attention is quadratic or linearized. Sub-quadratic variants such as linear attention~\cite{katharopoulos2020transformers} and sliding-window attention~\cite{beltagy2020longformer} reduce per-layer compute to $\mathcal{O}(N)$, but they still (i) propagate $N$-sized token-level intermediate tensors across depth, leaving memory at $\mathcal{O}(L \cdot N)$ rather than $\mathcal{O}(L \cdot K)$, and (ii) devote their interaction capacity to user-side token mixing, leaving user--item interaction capacity at $\mathcal{O}(L \cdot N)$---a factor of $K$ below SlimPer. We therefore expect SlimPer to retain a quality advantage over efficient-attention baselines, not merely an efficiency one, because the gain stems from \emph{how} capacity is allocated rather than from lowering the asymptotic order of attention.

\textit{Iterative access prevents irreversible loss.}
Standard bottleneck architectures are constrained by the Data Processing Inequality: for a Markov chain $X \!\to\! T \!\to\! Y$, information discarded during compression is irrecoverable. SlimPer breaks this chain---each layer cross-attends directly into the raw input tokens $X$, not only the compressed representation from the previous layer. In the language of the Information Bottleneck framework~\cite{tishby2000information}, SlimPer performs \textit{iterative} compression: each layer extracts additional task-relevant mutual information $I(\mathcal{X}^{k}; Y)$ while the raw evidence remains available to correct earlier omissions---a property validated by the attention heatmaps in Appendix~\ref{app:interpretability}, which show that successive layers progressively narrow their focus from broad historical context to the most relevant recent signals.

\section{Related Work}

We situate SlimPer within three lines of research: feature interaction learning for fixed-size inputs, user history modeling for variable-length event sequences, and cross-attention bottleneck architectures from adjacent domains.

\subsection{Feature Interaction Learning}

This paradigm focuses on capturing complex, high-order interactions among fixed-size input features. Factorization Machines~\cite{rendle2010fm} and DeepFM~\cite{guo2017deepfm} introduced efficient second-order interaction modeling through pairwise factorization. DLRM~\cite{naumov2019dlrm} combined sparse embeddings with MLP-processed dense features and explicit dot-product interactions, a design that has since been widely cited. Deep Cross Network (DCN)~\cite{wang2017dcn} and DCNv2~\cite{wang2020dcnv2} proposed cross networks to model bounded-degree feature crosses, while AutoInt~\cite{song2019autoint} applied multi-head self-attention over feature embeddings to learn implicit higher-order interactions. More recent work emphasizes architectural scalability: DHEN~\cite{zhang2022dhen} and Wukong~\cite{zhang2024wukong} systematically studied stacking interaction modules, demonstrating favorable scaling laws for fixed-size features.

\textit{Relation to SlimPer.} These approaches treat user history as a pre-aggregated feature vector. SlimPer subsumes their capability through its knowledge base refinement while additionally enabling direct access to individual user history events at every layer.

\subsection{User History Modeling}

Effectively capturing signals from sequence features has become increasingly important as user history data grows richer. DIN~\cite{zhou2018din} and DIEN~\cite{zhou2019dien} introduced target-aware attention over user interaction histories for intent modeling. Transformer-based models (BST~\cite{chen2019bst}, BERT4Rec~\cite{sun2019bert4rec}, SASRec~\cite{kang2018sasrec}) treat user events as tokens and apply full self-attention, achieving expressive user representations but incurring $\mathcal{O}(N^2)$ cost that limits scaling.

Research on ultra-long user histories has explored several strategies: \textit{(1) Retrieval-based}: SIM~\cite{pi2020sim} uses a two-stage hard search to retrieve relevant behaviors; \textit{(2) Memory-based}: MIMN~\cite{pi2019mimn} maintains fixed-size representations of long-term interests; \textit{(3) Compression-based}: LONGER~\cite{chai2025longerscalinglongsequence} applies token merging, and HSTU~\cite{zhai2024hstu} uses stochastic length sampling; \textit{(4) Offline aggregation}: TWIN-V2~\cite{si2024twinv2} and DV365~\cite{lyu2025dv365} shift long-range modeling to pre-computed embeddings. A common pattern across these strategies is that they reduce the cost of long histories by limiting how much raw behavioral detail is available for modeling user-item interactions, trading fine-grained, target-aware access to user history information for tractability at scale.

Despite these advances, user history modeling typically remains a separate component whose output is summarized and combined with fixed-size features via late fusion, limiting deep cross-modality interactions. Recent unified architectures, such as OneTrans~\cite{zhang2025onetrans}, HHFT~\cite{yu2025hhft}, RankMixer~\cite{zhu2025rankmixer}, process all modalities jointly using Transformer-style designs, but inherit the same quadratic cost when scaling to long sequences.

\textit{Relation to SlimPer.} SlimPer resolves this trade-off with $\mathcal{O}(K \cdot N)$ per-layer compute and a fixed-size memory footprint that does not grow with sequence length, scaling to 10k+ events while accessing raw user-side tokens at every layer without information loss.

\subsection{Cross-Attention Bottleneck Architectures}

SlimPer's use of a fixed-size knowledge base that iteratively cross-attends into a larger token set shares structural resemblance with architectures from other domains. Perceiver~\cite{jaegle2021perceiver} and Perceiver IO~\cite{jaegle2022perceiverio} maintain a learned latent array that iteratively cross-attends into raw inputs, achieving linear complexity in input length. Set Transformer~\cite{lee2019set} introduces Induced Set Attention Blocks (ISAB) with a similar bottleneck for set-structured data. Flamingo~\cite{alayrac2022flamingo} applies cross-attention from a language model into visual tokens for multi-modal understanding.

While SlimPer shares the principle of decoupling computation from input length through a compact bottleneck, it introduces several recommendation-specific inductive biases that are absent from general-purpose architectures: (1)~\textbf{discriminative relevance focus}: the knowledge base is initialized from item-side features and dedicated entirely to $\textit{<user, item>}$ relevance scoring; (2)~\textbf{explicit relevance matching}: a dedicated dot-product step (Step~2) computes scores between retrieved evidence and multifaceted templates, with no analog in Perceiver-style designs; (3)~\textbf{ROO awareness}: user-side tokens are computed once and shared across all candidates within a request, reducing redundant computation and memory usage; and (4)~\textbf{modality-dependent attention kernels}: SlimPer employs parametric MLP-based attention for fixed-size sparse tokens and scaled dot-product attention for variable-length sequence features.

\section{Experiments}

\subsection{Experiment Setup}

\paragraph{Dataset}

We conduct experiments on Instagram's final-stage ranking models using the same training and evaluation data used to train the current models.

\paragraph{Evaluation Metrics}

We evaluate model performance using Normalized Entropy (NE), a standard offline metric that exhibits directional alignment with A/B test results. NE normalizes the model's binary cross-entropy loss (BCEloss) by dividing it by the BCEloss of a naive baseline that predicts the global label average $p_k$ for all samples. An NE improvement of approximately 0.03\% is considered empirically significant. 

\begin{equation}
\text{NE}_k = \frac{\text{BCEloss}(\hat{y}^{(k)}, y^{(k)})}{-\left[p_k\ln(p_k) + (1-p_k)\ln(1-p_k)\right]},
\label{eq:normalized_entropy}
\end{equation}

\paragraph{Baseline}

We compare SlimPer against Instagram's existing models, which employ a late-fusion architecture that stitches together HSTU~\cite{zhai2024hstu} for modeling variable-length user interaction history sequences and Wukong~\cite{zhang2024wukong} for modeling fixed-size sparse/dense features. These models perform the final-stage ranking for Instagram's major surfaces and represent an established baseline. To ensure a fair comparison, we report NE improvements relative to the baseline using the same training and evaluation date ranges. All other experimental settings are held constant across models, including optimizer configuration and feature sets, unless otherwise noted.

\subsection{Offline Performance}

\label{sec:offline_performance}

We conducted an offline evaluation of SlimPer against the baseline on both Reels and Feed. The main results are summarized in Table~\ref{tab:ne_comp_all}. We make three observations that are consistent across both surfaces: (1) SlimPer achieves consistent NE improvements across all tasks. Due to space constraints, we report only representative tasks for both surfaces. (2) At the same sequence length, SlimPer delivers both NE wins and QPS wins compared to the baseline, indicating that the proposed design improves prediction quality while also improving efficiency. (3) We observe a scaling trend: NE wins increase as sequence length grows. For example, on Feed reshare prediction, NE improvements grow from -0.49\% at 1k sequence length to -0.94\% at 4k sequence length, corresponding to close to 2$\times$ relative gain in model quality, albeit with a 16\% QPS regression. 

These empirical findings are consistent with the Big-O analysis in Table~\ref{tab:complexity}: SlimPer's linear-in-$N$ computational complexity and its $\mathcal{O}(L \cdot K \cdot d)$ memory footprint directly account for the observed 11\%+ QPS improvement and 9--18\% memory reduction compared to the quadratic-cost baseline. Because this cost reduction stems from replacing self-attention over the full sequence with fixed-size knowledge base refinement, it generalizes to any transformer-like method that processes the full token sequence via quadratic self-attention.

\begin{table*}[t]
\centering
\small
\caption{SlimPer vs.\ baseline. NE and QPS are relative changes (\%); Memory is absolute change in average usage (\%); all w.r.t.\ surface-specific baselines. Lower NE/Memory and higher QPS are better.}
\label{tab:ne_comp_all}
\begin{tabular}{llcccccc cc}
\toprule
\multirow{2}{*}{\textbf{Surface}} &
\multirow{2}{*}{\textbf{Model Configuration}} &
\multicolumn{6}{c}{\textbf{NE (\%)}} &
\multirow{2}{*}{\textbf{QPS (\%)}} &
\multirow{2}{*}{\textbf{Memory Diff (\%)}} \\
\cmidrule(lr){3-8}
 & &
\textbf{reshare} & \textbf{skip} & \textbf{like} & \textbf{comment} & \textbf{follow} & \textbf{save} \\
\midrule
\multirow{3}{*}{\textbf{Reels}}
 & Baseline w/ \textbf{2k events} & 0.00 & 0.00 & 0.00 & 0.00 & 0.00 & 0.00 & 0.00 & 0.00 \\
 & SlimPer w/ \textbf{2k events} & \textbf{-0.51} & -0.31 & -0.23 & -0.41 & -0.35 & -0.35 & \textbf{+11.0} & \textbf{-9.32}\\
 & SlimPer w/ 5k events & \textbf{-0.80} & -0.57 & -0.49 & -0.64 & -0.53 & -0.62 & -10.0 & +4.59 \\
\midrule
\multirow{3}{*}{\textbf{Feed}}
 & Baseline w/ \textbf{1k events} & 0.00 & 0.00 & 0.00 & 0.00 & 0.00 & 0.00 & 0.00 & 0.00 \\
 & SlimPer w/ \textbf{1k events} & -0.49 & -0.28 & \textbf{-0.54} & -0.52 & -0.71 & -0.62 & \textbf{+12.5} & \textbf{-18.12} \\
 & SlimPer w/ 4k events & -0.94 & -0.52 & -0.91 & -0.94 & -1.00 & \textbf{-1.05} & -16.07 & +2.04 \\
\bottomrule
\end{tabular}
\end{table*}

\subsection{A/B Test Performance}

We launched SlimPer w/ \textbf{5k events} for Instagram Reels and SlimPer w/ \textbf{4k events} for Instagram Feed, the models described in Section~\ref{sec:offline_performance}. After pre-testing and backtesting, SlimPer was launched to full traffic on both surfaces, where it delivers statistically significant gains across multiple major engagement metrics, with an aggregate topline impact roughly 10$\times$ that of a typical statistically significant launch---a notable improvement from modeling changes in 2025. Moreover, SlimPer introduces no regression on ecosystem guardrail metrics (e.g., integrity, diversity, and recency) and maintains training and inference GPU capacity at approximately neutral levels, confirming that the efficiency gains observed offline (Section~\ref{sec:offline_performance}) translate to real-world serving.

In 2026, we have launched additional SlimPer variants that scale to 10k+ events with increased model capacity. These launches have delivered major engagement wins across multiple product surfaces at Meta, further validating SlimPer's efficacy and scalability under real-world constraints.

\subsection{Scaling and Efficiency Analysis}

\begin{figure}[t]
\centering
\includegraphics[width=\linewidth]{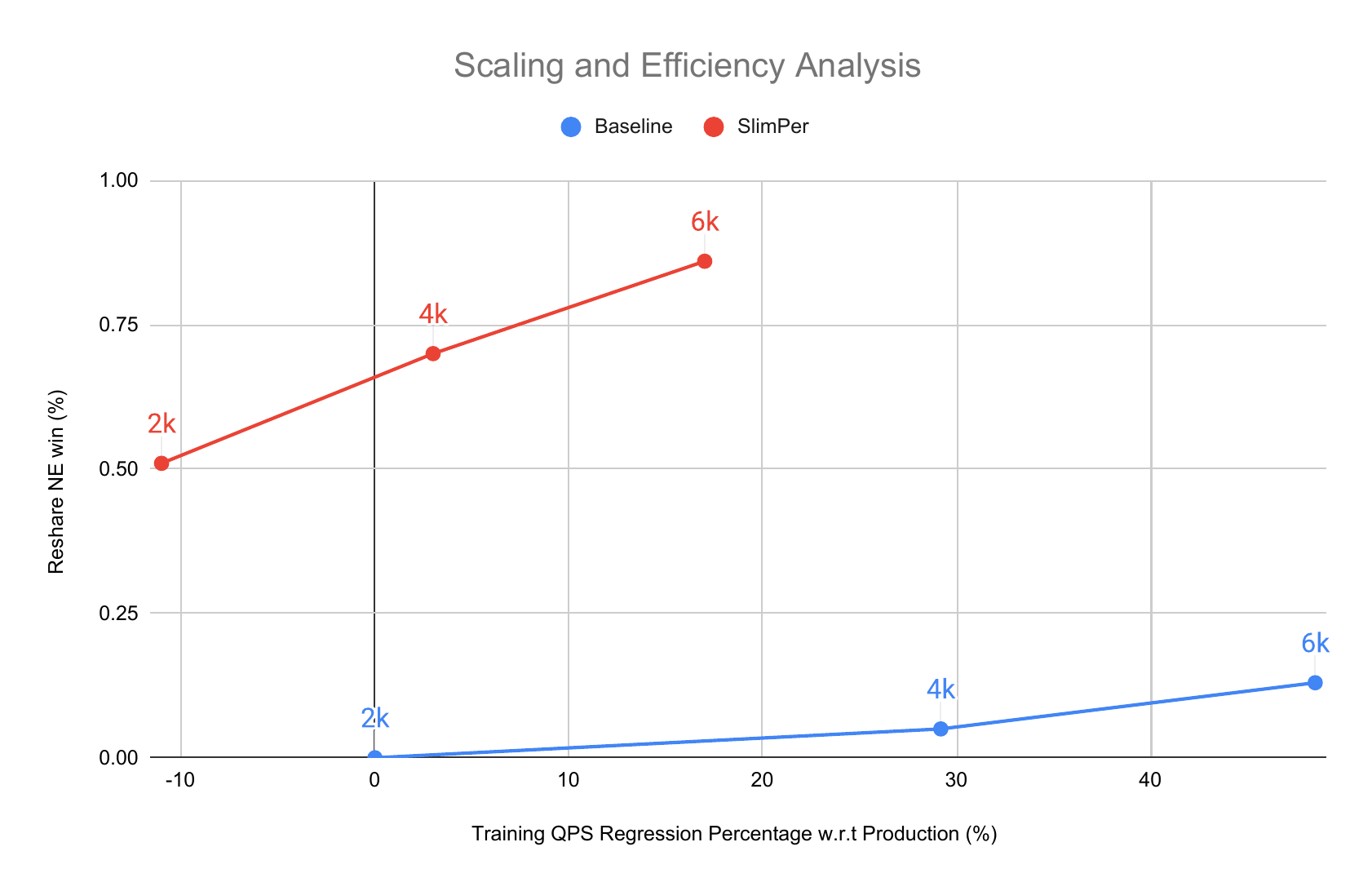}
\caption{Quality--efficiency scaling as user-history sequence length grows (2k--6k). Each curve traces one model (SlimPer vs.\ the baseline) across sequence lengths; the $x$-axis is training QPS regression and the $y$-axis is NE win, both relative to the baseline (\%). SlimPer sustains larger NE wins at smaller QPS regressions, and its advantage widens as sequence length increases.}
\label{fig:scaling}
\end{figure}

The Big-O analysis in Table~\ref{tab:complexity} predicts that SlimPer's $\mathcal{O}(K \cdot N)$ per-layer complexity should yield clear efficiency advantages over quadratic baselines, with the gap widening as sequence length grows. We validate this prediction empirically by comparing SlimPer against the baseline (late-fusion HSTU + Wukong) across different sequence length configurations, using the Reels model as a representative example. Whereas the baseline's usable history length is constrained by the quadratic attention cost over the sequence, SlimPer avoids this bottleneck. Results are summarized in Figure~\ref{fig:scaling}.

\textbf{Main findings.} At matched 2k length, SlimPer outperforms the baseline by $-0.51\%$ NE (reshare) while also delivering higher throughput ($+11\%$ QPS) and lower memory---improving quality and efficiency simultaneously. The gain compounds with sequence length: at 6k, SlimPer reaches $-0.86\%$ NE at under $20\%$ QPS regression, whereas the baseline gains only $-0.13\%$ NE while losing $48\%$ QPS. As the curves show, SlimPer's NE improves monotonically as the compact knowledge base absorbs richer long-term signals, while the quadratic-cost baseline sees diminishing returns at longer sequences.

\textbf{These results confirm that SlimPer's quality gains come not from spending more compute, but from allocating it more effectively toward user-item relevance matching.} Consistent with Table~\ref{tab:complexity} and the information-theoretic analysis in Section~\ref{sec:stacking}, SlimPer's $\mathcal{O}(K \cdot N)$ per-layer cost translates to $8$--$25\times$ fewer FLOPs in practice (Appendix~\ref{app:flops}), with the advantage growing superlinearly in input length: the compact knowledge base redirects interaction capacity away from the $\mathcal{O}(N^2)$ user-side-to-user-side computations that dominate transformer-like models and toward $\mathcal{O}(K \times N)$ relevance-focused interactions per layer.

\paragraph{Memory efficiency.} The observed 9--18\% memory reduction at matched sequence length stems from two complementary mechanisms: the fixed-size knowledge base replaces $\mathcal{O}(N)$-sized intermediate tensors at every layer, and the ROO-aware design shares a single copy of all user-side tokens across candidates within each request.
\subsection{Ablation Studies} 
We conduct ablation experiments to validate our design decisions.

\subsubsection{Effect of Contextual Encoding}
\label{sec:contextual_effect}

We ablate whether incorporating surrounding event information into each history token improves performance by comparing three strategies: (1)~no context (independent encoding), (2)~event-type context (encoding surrounding actions, e.g., like, reshare, long watch), and (3)~full context (event-type + event content jointly). Introducing contextual encoding yields approximately $-0.15\%$ NE improvement on the reshare task. Notably, the gain comes primarily from encoding event type alone; adding event content provides only marginal additional benefit ($<0.02\%$ NE). This suggests that users' behavioral patterns---\textit{what actions} they take---are more informative than the specific items they interact with, highlighting an important difference from language modeling where content dependencies determine meaning.

\subsubsection{Effect of Knowledge Base Size and Model Depth}

We further ablate the knowledge base size $K$ and the number of stacked SlimPer layers. A knowledge base size of $K$=64 provides a favorable quality-efficiency balance; reducing to $K$=4 yields +23.2\% QPS but degrades NE by +1.2\%, confirming the knowledge base's role as an effective information bottleneck. For model depth, 7 layers offers strong performance; gains continue up to 9 layers ($-0.12\%$ NE) with moderate cost increase. Full ablation tables are provided in Appendix~\ref{app:ablation}.

\subsection{Interpretability of Model Predictions}

SlimPer's attention mechanism enables direct attribution of recommendations to specific user history events. For a given request, the model produces attention scores between candidate items and sequence tokens, allowing practitioners to identify the top-$k$ historical interactions that most influenced each prediction. Spot-check analyses confirm that attention weights align with ranking outputs, indicating high matching fidelity. We provide a detailed analysis of attention patterns across model layers in Appendix~\ref{app:interpretability}.

For reproducibility details including hyperparameter settings, see Appendix~\ref{app:reproducibility}.

\section{Limitations}

\paragraph{Limited direct comparison with recent unified architectures.} We compare SlimPer against a mature baseline (HSTU + Wukong) rather than directly against recent transformer-style unified architectures (e.g., OneTrans, HHFT, RankMixer). In practice, a standalone unified HSTU performs worse than this hybrid, which uses dedicated components per modality (HSTU for user history, Wukong for fixed-size features); the hybrid is therefore a stronger comparison point than any single unified architecture, and SlimPer outperforms it by a clear margin while also simplifying the system into a single backbone. Differences in training data, feature sets, and serving constraints across organizations further make reported numbers from other systems non-comparable. Our complexity analysis (Table~\ref{tab:complexity}) is nonetheless general: any architecture using quadratic self-attention over the full sequence faces the same $\mathcal{O}(N^2)$ compute bottleneck, and any architecture maintaining $N$-sized intermediate tensors or duplicating user-side representations per candidate faces the same memory bottleneck---both of which SlimPer's fixed-size, ROO-aware knowledge base resolves.

\paragraph{Reproducibility on internal data.} All experiments are conducted on internal Instagram training and evaluation data, which limits independent reproduction of our exact results. Furthermore, our models are trained to convergence on data volumes not readily available in public benchmarks, which further limits exact reproducibility but also validates SlimPer's effectiveness in deployment. To mitigate this, we provide detailed architectural hyperparameters (Appendix~\ref{app:reproducibility}), describe the model using only standard building blocks (MLPs, QKV attention, linear projections, RMSNorm), and report relative improvements against a clearly defined baseline under controlled experimental conditions. We believe these details are sufficient for practitioners to reimplement and validate the approach on their own recommendation datasets.

\section{Conclusion}

We presented SlimPer, a unified personalization model that reframes recommendation ranking as iterative refinement of a compact $\textit{<user, item>}$ knowledge base. By departing from the transformer-mimicking paradigm dominant in recent work, SlimPer achieves a distinct efficiency-quality trade-off: model depth is decoupled from input sequence length, raw multi-modal evidence is preserved uncompressed and accessed directly at every layer, and the entire model capacity is focused on relevance estimation rather than maintaining token-level representations. SlimPer further minimizes GPU memory by eliminating $N$-sized intermediate tensors across layers and maintaining a single shared copy of user-side tokens across all candidates within a request via request-only optimization.

Our experiments on an industrial recommender system validate the effectiveness of SlimPer. The proposed architecture achieves strong offline performance with lower computational cost than transformer-based alternatives, and delivers statistically significant engagement improvements in controlled A/B experiments. Moreover, SlimPer is inherently interpretable: attention weights at each refinement stage reveal how evidence is selected and aggregated, facilitating principled debugging, auditing, and model analysis.

Looking ahead, we see several promising directions for future work. First, the iterative refinement paradigm naturally extends beyond sequential user histories to evidence sources without inherent ordering, such as cross-surface engagement signals or candidates generated by upstream selection systems. Because the knowledge base interacts directly with raw tokens rather than relying on sequence structure, SlimPer can integrate such signals without architectural changes. Second, SlimPer's whole-model ROO-aware design enables efficient deployment across ranking stages, such as early-stage ranking, where larger candidate sets further increase the benefits of amortized computation. Finally, the current matching and refinement modules are intentionally lightweight, relying on simple dot-product matching and MLP-based updates. Exploring more expressive operators within these components may yield additional quality improvements while preserving the iterative-refinement framework with its $\mathcal{O}(N)$ per-layer complexity and fixed-size memory footprint. More broadly, we believe that recommendation-centric architectural design around the unique computational and modeling requirements of personalization will continue to drive practical advances in industrial recommendation systems.

\newpage
\begin{acks}
\paragraph{Engineering Collaborators} We are grateful to the many collaborators across Meta whose partnership has been invaluable in building and scaling SlimPer across our family of apps: Jay Wang, Keming Zhang, Chen Chen, Qichao Que, Wanli Ma, Maxwell Lin-He, Yu Liu, Yanhong Wu, Xinye Zheng, Wen-Yun Yang, Rui Zhang, Jiyan Yang, Zichao Li, Lei Nie, Yan Sun, Rongcheng Lin, Ivan Ji, Cory Nezin, Jeffrey Cordero, Lambert Chu, Leon Yan, Keyi Wu, Yiming Liao, Irra Na, Neel Pawar, Tingting Zhang, Xue Zou, Ziwei Li, Zhe (Joe) Wang, Haifeng Lan, Qinjie Lin, Ye Wang, Huasha Zhao, Jianming He, William Pei, Dawei Fan, Yezhou Huang, Minhan Li, Zhiwei Zhang, Dev (Devashish) Shankar, Jackie (Jiaqi) Xu, Chunzhi Yang, Haoxun Luo, Hao Lin, Prashasti Baid, Lu Zhang, Liang Guo, and Rui Jian.
\paragraph{Leadership} We are also grateful for the leadership support spanning Meta's family of apps and MRS: Jayant Subramanian, Haotian Wu, Jing Qian, Hui Zhang, Shilin Ding, Sophia (Xueyao) Liang, Xinyao Hu, Linhong Zhu, Yisong Song, Shujian Bu, Aameek Singh, Hong Yan, Meihong Wang, Mahesh Srinivasan, Adam (Yang) Song, Nipun Mathur, Lars Backstrom, Jeff Huang, Tessa Lyons-Laing, Deepak Agarwal, and Max Eulenstein, whose guidance and investment made this work possible.
\end{acks}

\appendix \label{app:extra-exp}
\section{Reproducibility} \label{app:reproducibility}

Table~\ref{tab:hyperparams} summarizes the main architectural hyperparameters for SlimPer on Feed and Reels. All models are trained for one epoch with the Shampoo optimizer~\cite{gupta2018shampoo} at a learning rate of 0.05, with identical optimizer settings across all model configurations (both SlimPer and baseline).

\begin{table}[H]
\centering
\small
\caption{SlimPer architectural hyperparameters for Feed and Reels.}
\label{tab:hyperparams}
\begin{tabular}{lcc}
\toprule
\textbf{Hyperparameter} & \textbf{Feed} & \textbf{Reels} \\
\midrule
Number of SlimPer layers ($L$)       & 5 & 7 \\
Knowledge base size ($K$)           & 64 & 64 \\
Hidden dimension ($d$)               & 256 & 256 \\
Query slots ($q$)         & 16 & 16 \\
Template slots ($t$)      & 32 & 32 \\
\bottomrule
\end{tabular}
\end{table}

\section{Notation and Preliminaries} \label{app:notation}

To reduce architectural complexity and simplify the implementation, we restrict the model to a minimal set of building blocks with identical internal structures, differing only in hyperparameter settings. We define the following notations and operator blocks used throughout the network. 

\paragraph{Multilayer Perceptrons ($\mathrm{MLP}$)}

$\mathrm{MLP}$ denotes the standard multilayer perceptron, a stack of fully connected layers with nonlinear activations applied between layers. In our experiments, we use SiLU as the activation function.

\paragraph{Segment-wise Linear Projection ($\mathcal{L}$)}

$\mathcal{L}$ is a segment-wise linear projection module that transforms an input tensor $\alpha \in \mathbb{R}^{L_{\mathrm{in}}  \times d}$ into an output tensor  $\beta \in \mathbb{R}^{L_{\mathrm{out}} \times d}$ using a learnable projection matrix $\rho$:

\begin{equation}
\beta = \mathcal{L}(\alpha) = \rho \alpha,
\quad \rho \in \mathbb{R}^{L_{\mathrm{out}} \times L_{\mathrm{in}}}.
\end{equation}

In our network, we use this module multiple times, each with independent learnable parameters. For simplicity, we omit the distinguishing subscripts and superscripts in subsequent equations.

\paragraph{All-Modality-Aware $\textit{<user, item>}$ Knowledge Base ($\mathcal{X}$)}

We use $\mathcal{X}$ to denote the all-modality $\textit{<user, item>}$ knowledge base. This knowledge base is initialized at Layer-0 as:

\begin{equation}
    \mathcal{X}^{0} = \mathcal{L}(\mathbf{S}_{\text{in}})
\end{equation}

where $\mathbf{S}_{\text{in}}$ represents the non-user-side input tokens, typically derived from item-side or cross-side sparse features. The shape of $\mathcal{X}^{k}$ can vary across layers, but in our experiments, we use a consistent shape of $K \times d$ (with $K=64$) for all layers.
\section{Tokenization Details} \label{app:tokenization}

All raw input features are first converted into a set of modality-specific tokens within the unified backbone. This step preserves the semantic distinctions of the feature types while preparing them for the shared interaction stack. We consider three types of input modalities: sparse features, user history sequence features and dense features. The corresponding tokenization approaches are summarized in Table~\ref{tab:continuous_tokenization}.

\begin{table}[t]
    \centering
    \footnotesize
    \caption{Continuous tokenization approaches across different information modalities.}
    \label{tab:continuous_tokenization}
    \setlength{\tabcolsep}{6pt}
    \renewcommand{\arraystretch}{1.1}
    \begin{tabular}{@{}p{0.11\textwidth} p{0.34\textwidth}@{}}
        \toprule
        \textbf{Modality} & \textbf{Continuous Tokenization Approaches} \\
        \midrule
        Sparse &
        \textbf{Embedding Pooling (E-P):} Sparse features are tokenized via sum or mean pooling over embeddings, yielding one or a small number of fixed-size vectors (e.g., $1 \times d$). \\
        \addlinespace
        Sequence &
        \textbf{Event Modeling (E-M):} All information describing each event is aggregated and fused into a fixed-length event representation (e.g., $1 \times d$ per event). \\
        Dense &
        \textbf{Dense Processing (D-P):} Dense features are normalized and projected by an MLP into a fixed-size vector (e.g., 1024). \\
        \addlinespace
        \bottomrule
    \end{tabular}
\end{table}

\subsubsection{Multi-modality Tokenization}

\paragraph{Embedding-Pooling Module (E-P)}

Sparse features are first mapped to embedding vectors through standard embedding lookup tables. For features with multiple values (e.g., a user's top-$k$ liked categories), we apply pooling operations:

\begin{equation}
    \mathbf{S}^{j} = \operatorname{Pool}\left(\{\mathbf{e}_{j,1}, \mathbf{e}_{j,2}, \ldots, \mathbf{e}_{j,k}\}\right)
\end{equation}

where $\operatorname{Pool}(\cdot)$ can be mean or sum pooling. Concatenating the pooled results from all sparse features, we obtain the tensor representing sparse feature tokens as $\mathbf{S} = (\mathbf{S}^{(1)}, \ldots, \mathbf{S}^{(|\mathbf{S}|)}) \in \mathbb{R}^{|S| \times d}$, with $d$ as the dimension for each token.

\paragraph{Event-Modeling Module (E-M)}

The E-M module processes each event in the user history into a token representation. Each event is typically described by several sparse features representing its content, the user's explicit action types, implicit watch time, engagement timestamps, and other metadata. We use $\mathbf{E} = (\mathbf{E}^{(1)}, \ldots, \mathbf{E}^{(t)}, \ldots)$ to represent the tokenized user history event sequence. To obtain each $\mathbf{E}^{(t)}$, the E-M module builds several parallel encoding components, projects each to the shared embedding dimension $d$, and combines them via summation:

\begin{equation}
\begin{aligned}
\mathbf{E}^{(t)} = \;
    & \mathrm{MLP}_{E}^{s}(\mathbf{e}_{\text{sparse}}^{(t)}) + \mathbf{e}_{\text{temporal}}^{(t)} \\
    & + \mathrm{MLP}_{E}^{e}(\mathbf{e}_{\text{type}}^{(t)}) + \mathbf{e}_{\text{context}}^{(t)}
\end{aligned}
\end{equation}

where $\mathbf{e}_{\text{sparse}}^{(t)}$ is the content embedding from concatenated pooled sparse features describing the event item, $\mathbf{e}_{\text{type}}^{(t)}$ is the event type encoding that captures both explicit and implicit actions, and $\mathbf{e}_{\text{temporal}}^{(t)}$ is the temporal encoding, which is the sum of positional encoding and delta-time encoding.

\paragraph{Dense-Processing Module (D-P)}

Dense features are concatenated and projected into a fixed-size vector through a simple MLP:

\begin{equation}
\mathbf{D} =
\mathrm{MLP}
\left(
\operatorname{Concat}(\mathbf{d}_1, \mathbf{d}_2, \ldots, \mathbf{d}_{|\mathcal{D}|})
\right)
\end{equation}

\section{Ablation Studies} \label{app:ablation}

We conduct ablation experiments to validate our design decisions underlying SlimPer's iterative refinement paradigm. The baseline corresponds to the \textbf{Reels SlimPer model with 5k events} used in Table~\ref{tab:ne_comp_all} with knowledge base size $K=64$ and 7 refinement layers. 

\subsubsection{Effect of Knowledge Base Size}

Table~\ref{tab:kb_size} presents an ablation study on the number of knowledge base slots $K$ (i.e., $K \in \{4, 8, 16, 32, 64\}$), which directly evaluates the information capacity of the iterative refinement bottleneck. The baseline configuration has $K=64$. For each configuration, the query size $q$ is defined as one quarter of the knowledge base size, while the template size $t$ is defined as one half of $K$.

\begin{table}[H]
\centering
\small
\caption{Ablation study on knowledge base size (relative to the $K=64$ baseline).}
\begin{tabular}{cccc}
\toprule
\textbf{K, q, t} & \textbf{NE (\%)} & \textbf{QPS (\%)} & \textbf{Mem Diff (\%)} \\
\midrule
(64, 16, 32) & 0.0 & 0.0 & 0.0 \\
\midrule
(4, 1, 2) & +1.2 & +23.2 & -9.53 \\
(8, 2, 4) & +0.53 & +21.2 & -8.19 \\
(16, 4, 8) & +0.36 & +17.8 & -7.43 \\
(32, 8, 16) & +0.18 & +4.25 & -1.31 \\
\bottomrule
\end{tabular}
\label{tab:kb_size}
\end{table}

\subsubsection{Effect of Number of Refinement Layers}

Table~\ref{tab:layer_ablation} presents an ablation study on the number of stacked SlimPer refinement layers (i.e., $L \in \{3, 5, 7, 9\}$). This experiment evaluates whether iterative refinement consistently improves ranking quality and examines the depth at which performance gains begin to saturate. The baseline configuration has 7 refinement layers.

\begin{table}[H]
\centering
\small
\caption{Ablation study on the number of SlimPer refinement layers (relative to the 7-layer baseline).}
\begin{tabular}{cccc}
\toprule
\textbf{Layer} & \textbf{NE (\%)} & \textbf{QPS (\%)} & \textbf{Mem Diff (\%)} \\
\midrule
7 & 0.0 & 0.0 & 0.0 \\
\midrule
3 & +0.32 & +21 & -8.60  \\
5 & +0.14 & +8.8 & -5.50  \\
9 & -0.12 & -7.16 & +4.92 \\
\bottomrule
\end{tabular}
\label{tab:layer_ablation}
\end{table}

\section{Interpretability of Model Predictions} \label{app:interpretability}

SlimPer's iterative refinement design provides intrinsic interpretability: the QKV attention at each layer directly links past user interactions to the current prediction, making it possible to attribute recommendations to specific historical events. Unlike prior models that stitch together opaque components, SlimPer's attention scores offer a natural explanation for why an item is recommended to a user.

\paragraph{Attention Analysis Across Layers.} For a given request, the model produces an attention matrix mapping candidate items (queries) to sequence tokens (keys). We aggregate attention distributions across 200 requests with 5k sequence length and visualize them as heatmaps, where the x-axis represents sequence indices in sequential position and the y-axis represents candidate items.
\begin{figure}[H]
  \centering
  \includegraphics[width=0.85\linewidth]{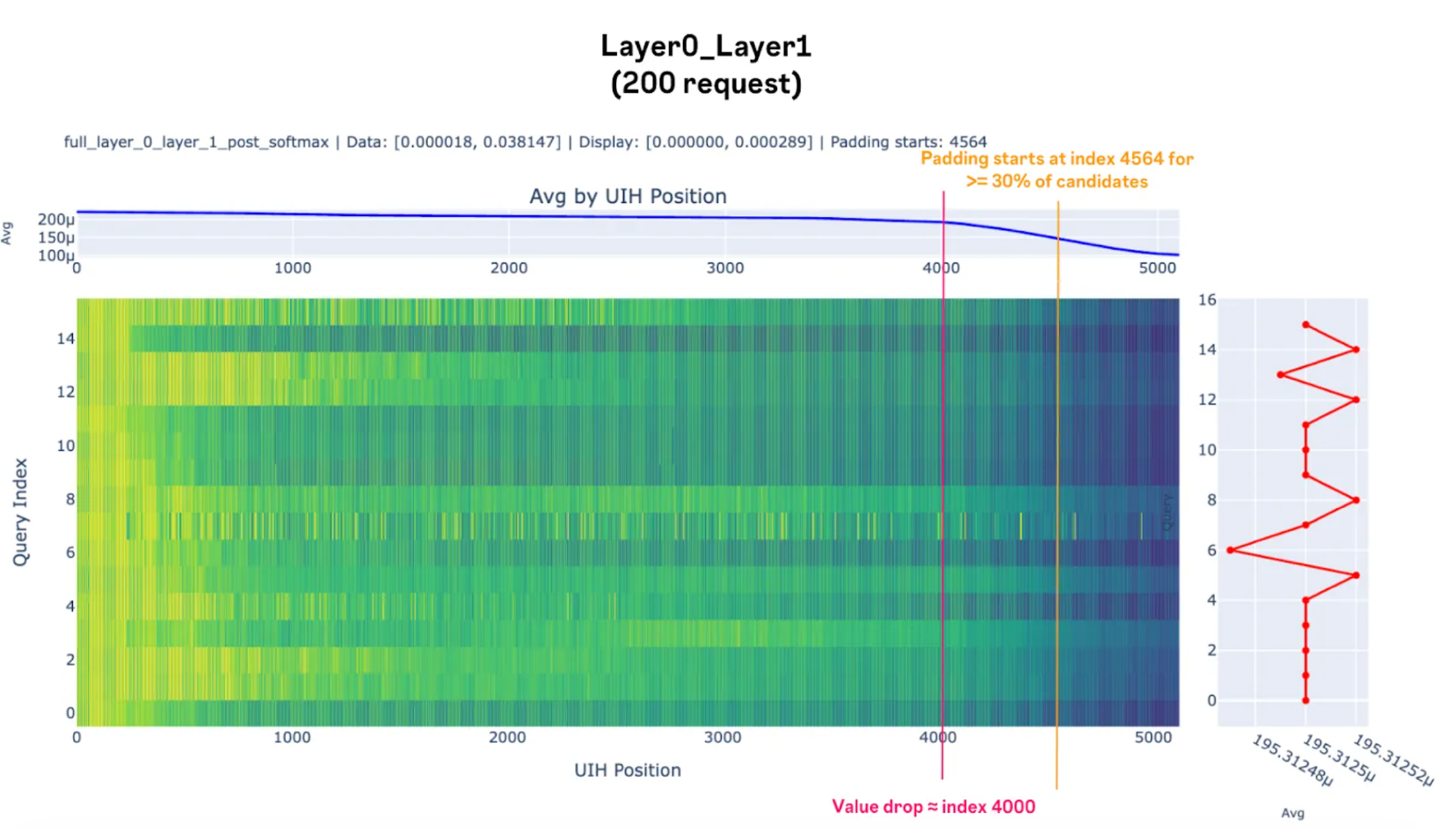}\par\vspace{1mm}
  \includegraphics[width=0.85\linewidth]{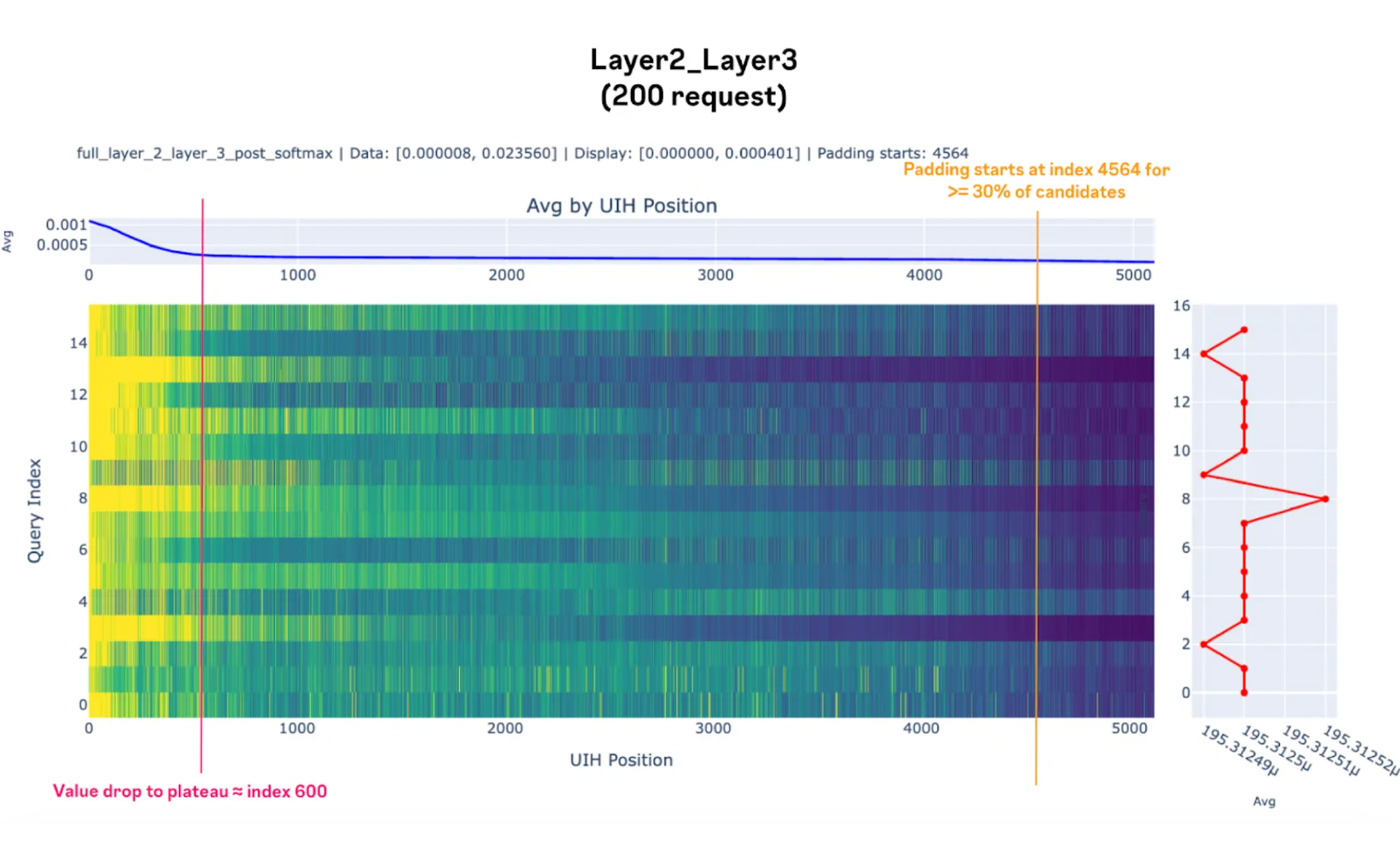}
  \caption{Attention heatmaps for lower and middle SlimPer layers. Lower layers attend broadly across the full history; middle layers concentrate on more recent events.}
  \label{fig:lower_layer}
\end{figure}

As shown in Fig.~\ref{fig:lower_layer}, at the lowest layers the model attends broadly across the first 4k sequence indices, capturing coarse-grained historical context. At the middle layers, attention concentrates on the first few hundred indices, focusing on a narrower and more recent subset of interactions.

\begin{figure}[t]
    \centering
    \includegraphics[width=0.85\linewidth]{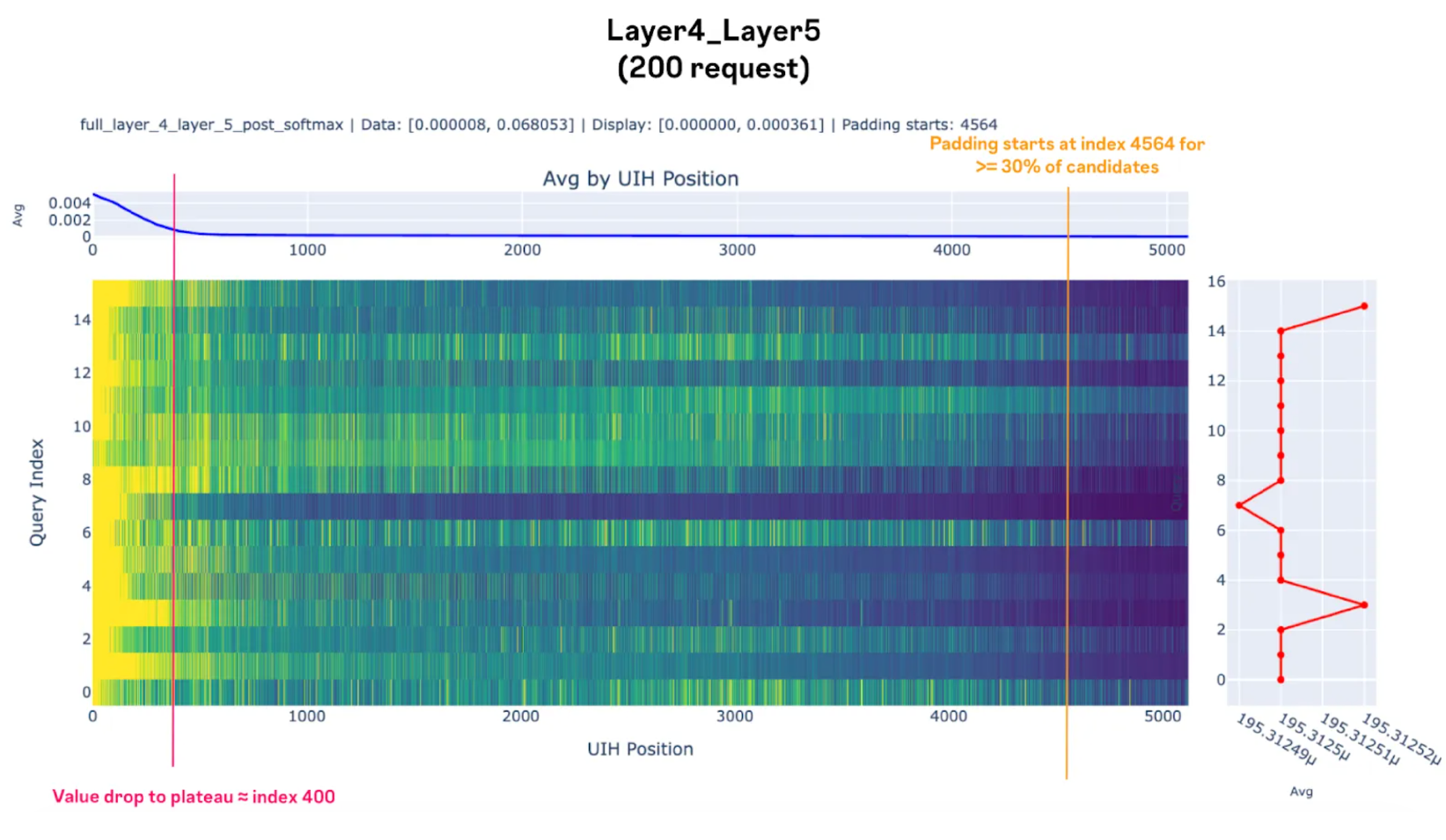}
    \caption{Attention heatmap for upper SlimPer layers. Attention further narrows to the most recent ${\sim}400$ events.}
    \label{fig:high_layer}
\end{figure}

At the upper layers (Fig.~\ref{fig:high_layer}), the pattern sharpens further---attention is concentrated within the most recent ${\sim}400$ events. This hierarchical attention behavior---broad context at lower layers, progressively narrower focus at upper layers---is consistent with SlimPer's iterative refinement design: earlier layers gather broad evidence about user preferences, while later layers refine predictions using the most recent and most influential behavioral signals.

\section{FLOPs Analysis} \label{app:flops}
Beyond throughput (QPS), we quantify the computational cost by counting all matrix multiplications in the forward pass (excluding embedding lookups, normalizations, and activations). Table~\ref{tab:flops} reports the estimated FLOPs at different sequence lengths. At the shared default of 2k, SlimPer uses ${\sim}8\times$ fewer FLOPs than the baseline. The gap widens with sequence length: at 6k, SlimPer (${\sim}6$~GFLOPs) still uses ${\sim}4\times$ fewer FLOPs than the baseline at its native 2k (${\sim}24$~GFLOPs), while achieving $-0.86\%$ NE improvement.

\begin{table}[H]
\centering
\small
\caption{Estimated forward-pass FLOPs per example (GFLOPs), counting all matrix multiplications. Baseline counts the HSTU encoder (the sequence-length-dependent component) only.}
\label{tab:flops}
\begin{tabular}{lccc}
\toprule
\textbf{Seq. Len.} & \textbf{Baseline (HSTU)} & \textbf{SlimPer} & \textbf{Ratio} \\
\midrule
2048 & ${\sim}24$ & ${\sim}3$ & ${\sim}8\times$ \\
4096 & ${\sim}74$ & ${\sim}5$ & ${\sim}16\times$ \\
6144 & ${\sim}150$ & ${\sim}6$ & ${\sim}25\times$ \\
\bottomrule
\end{tabular}
\end{table}

\bibliographystyle{ACM-Reference-Format}
\bibliography{refs}

@inproceedings{rendle2010fm,
  author={Rendle, Steffen},
  booktitle={2010 IEEE International Conference on Data Mining}, 
  title={Factorization Machines}, 
  year={2010},
  volume={},
  number={},
  pages={995-1000},
}

@inproceedings{wang2020dcnv2,
  title     = {DCN V2: Improved Deep \& Cross Network and Practical Lessons for Web-scale Learning to Rank Systems},
  author    = {Wang, Ruoxi and Shivanna, Rakesh and Cheng, Derek Zhiyuan and Jain, Sagar and Lin, Dong and Hong, Lichan and Chi, Ed H.},
  booktitle = {Proceedings of The Web Conference 2021},
  year      = {2021},
  pages     = {1785--1797},
}

@article{naumov2019dlrm,
  title={Deep learning recommendation model for personalization and recommendation systems},
  author={Naumov, Maxim and Mudigere, Dheevatsa and Shi, Hao-Jun Michael and Huang, Jianyu and Sundaraman, Narayanan and Park, Jongsoo and Wang, Xiaodong and Gupta, Udit and Wu, Carole-Jean and Azzolini, Alisson G and others},
  journal={arXiv preprint arXiv:1906.00091},
  year={2019}
}

@article{zhang2022dhen,
  title={DHEN: A deep and hierarchical ensemble network for large-scale click-through rate prediction},
  author={Zhang, Buyun and Luo, Liang and Liu, Xi and Li, Jay and Chen, Zeliang and Zhang, Weilin and Wei, Xiaohan and Hao, Yuchen and Tsang, Michael and Wang, Wenjun and others},
  journal={arXiv preprint arXiv:2203.11014},
  year={2022}
}

@InProceedings{zhang2024wukong,
  title = {Wukong: Towards a Scaling Law for Large-Scale Recommendation},
  author = {Zhang, Buyun and Luo, Liang and Chen, Yuxin and Nie, Jade and Liu, Xi and Li, Shen and Zhao, Yanli and Hao, Yuchen and Yao, Yantao and Wen, Ellie Dingqiao and Park, Jongsoo and Naumov, Maxim and Chen, Wenlin},
  booktitle = {Proceedings of the 41st International Conference on Machine Learning},
  pages = {59421--59434},
  year = {2024},
}

@inproceedings{zhou2018din,
  title={Deep interest network for click-through rate prediction},
  author={Zhou, Guorui and Zhu, Xiaoqiang and Song, Chenru and Fan, Ying and Zhu, Han and Ma, Xiao and Yan, Yanghui and Jin, Junqi and Li, Han and Gai, Kun},
  booktitle={Proceedings of the 24th ACM SIGKDD international conference on knowledge discovery \& data mining},
  pages={1059--1068},
  year={2018}
}

@inproceedings{zhou2019dien,
  title={Deep interest evolution network for click-through rate prediction},
  author={Zhou, Guorui and Mou, Na and Fan, Ying and Pi, Qi and Bian, Weijie and Zhou, Chang and Zhu, Xiaoqiang and Gai, Kun},
  booktitle={Proceedings of the AAAI conference on artificial intelligence},
  volume={33},
  number={01},
  pages={5941--5948},
  year={2019}
}

@inproceedings{si2024twinv2,
  title={Twin v2: Scaling ultra-long user behavior sequence modeling for enhanced ctr prediction at kuaishou},
  author={Si, Zihua and Guan, Lin and Sun, ZhongXiang and Zang, Xiaoxue and Lu, Jing and Hui, Yiqun and Cao, Xingchao and Yang, Zeyu and Zheng, Yichen and Leng, Dewei and others},
  booktitle={Proceedings of the 33rd ACM International Conference on Information and Knowledge Management},
  pages={4890--4897},
  year={2024}
}

@inproceedings{zhai2024hstu,
  title = {Actions Speak Louder than Words: Trillion-Parameter Sequential Transducers for Generative Recommendations},
  author = {Zhai, Jiaqi and Liao, Lucy and Liu, Xing and Wang, Yueming and Li, Rui and Cao, Xuan and Gao, Leon and Gong, Zhaojie and Gu, Fangda and He, Jiayuan and Lu, Yinghai and Shi, Yu},
  booktitle = {Proceedings of the 41st International Conference on Machine Learning},
  pages = {58484--58509},
  year = {2024},
}

@article{zhang2025onetrans,
  title={OneTrans: Unified Feature Interaction and Sequence Modeling with One Transformer in Industrial Recommender},
  author={Zhang, Zhaoqi and Pei, Haolei and Guo, Jun and Wang, Tianyu and Feng, Yufei and Sun, Hui and Liu, Shaowei and Sun, Aixin},
  journal={arXiv preprint arXiv:2510.26104},
  year={2025}
}

@article{yu2025hhft,
  title={HHFT: Hierarchical Heterogeneous Feature Transformer for Recommendation Systems},
  author={Yu, Liren and Zhang, Wenming and Zhou, Silu and Zhang, Tao and Zhang, Zhixuan and Ou, Dan},
  journal={arXiv preprint arXiv:2511.20235},
  year={2025}
}

@inproceedings{guo2017deepfm,
  title     = {DeepFM: A Factorization-Machine Based Neural Network for CTR Prediction},
  author    = {Guo, Huifeng and Tang, Ruiming and Ye, Yunming and Li, Zhenguo and He, Xiuqiang},
  booktitle = {Proceedings of the 26th International Joint Conference on Artificial Intelligence (IJCAI)},
  year      = {2017},
  pages     = {1725--1731}
}

@inproceedings{wang2017dcn,
  title     = {Deep \& Cross Network for Ad Click Predictions},
  author    = {Wang, Ruoxi and Fu, Bin and Fu, Gang and Wang, Mingliang},
  booktitle = {Proceedings of the 2017 ACM on Conference on Information and Knowledge Management},
  year      = {2017},
  pages     = {12:1--12:7},
}

@inproceedings{song2019autoint,
  title     = {AutoInt: Automatic Feature Interaction Learning via Self-Attentive Neural Networks},
  author    = {Song, Weiping and Shi, Chence and Xiao, Zhiping and Duan, Zhijian and Xu, Yewen and Zhang, Ming and Tang, Jian},
  booktitle = {Proceedings of the 28th ACM International Conference on Information and Knowledge Management (CIKM '19)},
  pages     = {1161--1170},
  year      = {2019},
}

@inproceedings{chen2019bst,
  title={Behavior sequence transformer for e-commerce recommendation in alibaba},
  author={Chen, Qiwei and Zhao, Huan and Li, Wei and Huang, Pipei and Ou, Wenwu},
  booktitle={Proceedings of the 1st international workshop on deep learning practice for high-dimensional sparse data},
  pages={1--4},
  year={2019}
}

@inproceedings{sun2019bert4rec,
  title={BERT4Rec: Sequential recommendation with bidirectional encoder representations from transformer},
  author={Sun, Fei and Liu, Jun and Wu, Jian and Pei, Changhua and Lin, Xiao and Ou, Wenwu and Jiang, Peng},
  booktitle={Proceedings of the 28th ACM international conference on information and knowledge management},
  pages={1441--1450},
  year={2019}
}

@inproceedings{kang2018sasrec,
  title={Self-attentive sequential recommendation},
  author={Kang, Wang-Cheng and McAuley, Julian},
  booktitle={2018 IEEE international conference on data mining (ICDM)},
  pages={197--206},
  year={2018},
  organization={IEEE}
}

@inproceedings{pi2020sim,
  title={Search-based user interest modeling with lifelong sequential behavior data for click-through rate prediction},
  author={Pi, Qi and Zhou, Guorui and Zhang, Yujing and Wang, Zhe and Ren, Lejian and Fan, Ying and Zhu, Xiaoqiang and Gai, Kun},
  booktitle={Proceedings of the 29th ACM International Conference on Information \& Knowledge Management},
  pages={2685--2692},
  year={2020}
}

@inproceedings{pi2019mimn,
  title={Practice on long sequential user behavior modeling for click-through rate prediction},
  author={Pi, Qi and Bian, Weijie and Zhou, Guorui and Zhu, Xiaoqiang and Gai, Kun},
  booktitle={Proceedings of the 25th ACM SIGKDD international conference on knowledge discovery \& data mining},
  pages={2671--2679},
  year={2019}
}

@inproceedings{lyu2025dv365,
  title     = {DV365: Extremely Long User History Modeling at Instagram},
  author    = {Lyu, Wenhan and Tyagi, Devashish and Yang, Yihang and Li, Ziwei and Somani, Ajay and Shanmugasundaram, Karthikeyan and Andrejevic, Nikola and Adeputra, Ferdi and Zeng, Curtis and Singh, Arun K. and Ransan, Maxime and Jain, Sagar},
  booktitle = {Proceedings of the 31st ACM SIGKDD Conference on Knowledge Discovery and Data Mining (KDD '25)},
  year      = {2025},
  pages     = {4717--4727},
}

@inproceedings{zhu2025rankmixer,
  title={Rankmixer: Scaling up ranking models in industrial recommenders},
  author={Zhu, Jie and Fan, Zhifang and Zhu, Xiaoxie and Jiang, Yuchen and Wang, Hangyu and Han, Xintian and Ding, Haoran and Wang, Xinmin and Zhao, Wenlin and Gong, Zhen and others},
  booktitle={Proceedings of the 34th ACM International Conference on Information and Knowledge Management},
  pages={6309--6316},
  year={2025}
}

@inproceedings{zeng2024interformer,
  title     = {InterFormer: Towards Effective Heterogeneous Interaction Learning for Click-Through Rate Prediction},
  author    = {Zeng, Zhichen and Liu, Xiaolong and Hang, Mengyue and Liu, Xiaoyi and Zhou, Qinghai and Yang, Chaofei and Liu, Yiqun and Ruan, Yichen and Chen, Laming and Chen, Yuxin and Hao, Yujia and Xu, Jiaqi and Nie, Jade and Liu, Xi and Zhang, Buyun and Wen, Wei and Yuan, Siyang and Wang, Kai and Chen, Wen-Yen and Han, Yiping and Li, Huayu and Yang, Chunzhi and Long, Bo and Yu, Philip S. and Tong, Hanghang and Yang, Jiyan},
  booktitle = {Proceedings of the 34th ACM International Conference on Information and Knowledge Management (CIKM '25)},
  pages     = {6225--6233},
  year      = {2025},
}

@article{lecun1998gradientbased,
  title={Gradient-based learning applied to document recognition},
  author={LeCun, Yann and Bottou, L{\'e}on and Bengio, Yoshua and Haffner, Patrick},
  journal={Proceedings of the IEEE},
  volume={86},
  number={11},
  pages={2278--2324},
  year={1998},
  publisher={IEEE}
}

@article{beltagy2020longformer,
  title={Longformer: The Long-Document Transformer},
  author={Beltagy, Iz and Peters, Matthew E and Cohan, Arman},
  journal={arXiv preprint arXiv:2004.05150},
  year={2020}
}

@inproceedings{vaswani2017attention,
  title={Attention is all you need},
  author={Vaswani, Ashish and Shazeer, Noam and Parmar, Niki and Uszkoreit, Jakob and Jones, Llion and Gomez, Aidan N and Kaiser, Lukasz and Polosukhin, Illia},
  booktitle={Advances in neural information processing systems},
  pages={5998--6008},
  year={2017},
  url={https://arxiv.org/abs/1706.03762}
}

@article{zhang2019rmsnorm,
  title={Root Mean Square Layer Normalization},
  author={Zhang, Biao and Sennrich, Rico},
  journal={arXiv preprint arXiv:1910.07467},
  year={2019}
}

@inproceedings{gupta2018shampoo,
  title={Shampoo: Preconditioned Stochastic Tensor Optimization},
  author={Gupta, Vineet and Koren, Tomer and Singer, Yoram},
  booktitle={International Conference on Machine Learning},
  pages={1842--1850},
  year={2018},
  organization={PMLR}
}

@inproceedings{jaegle2021perceiver,
  title={Perceiver: General Perception with Iterative Attention},
  author={Jaegle, Andrew and Gimeno, Felix and Brock, Andrew and Vinyals, Oriol and Zisserman, Andrew and Carreira, Joao},
  booktitle={International Conference on Machine Learning (ICML)},
  year={2021}
}

@inproceedings{jaegle2022perceiverio,
  title={Perceiver {IO}: A General Architecture for Structured Inputs \& Outputs},
  author={Jaegle, Andrew and Borgeaud, Sebastian and Alayrac, Jean-Baptiste and Doersch, Carl and Ionescu, Catalin and Ding, David and Koppula, Skanda and Zoran, Daniel and Brock, Andrew and Shelhamer, Evan and Hénaff, Olivier and Botvinick, Matthew M. and Zisserman, Andrew and Vinyals, Oriol and Carreira, João},
  booktitle={International Conference on Learning Representations (ICLR)},
  year={2022}
}

@inproceedings{lee2019set,
  title={Set Transformer: A Framework for Attention-based Permutation-Invariant Neural Networks},
  author={Lee, Juho and Lee, Yoonho and Kim, Jungtaek and Kosiorek, Adam and Choi, Seungjin and Teh, Yee Whye},
  booktitle={International Conference on Machine Learning (ICML)},
  year={2019}
}

@inproceedings{alayrac2022flamingo,
  title={Flamingo: a Visual Language Model for Few-Shot Learning},
  author={Alayrac, Jean-Baptiste and Donahue, Jeff and Luc, Pauline and Miech, Antoine and Barr, Iain and Hasson, Yana and Lenc, Karel and Mensch, Arthur and Millican, Katherine and Reynolds, Malcolm and Ring, Roman and Rutherford, Eliza and Cabi, Serkan and Han, Tengda and Gong, Zhitao and Samangooei, Sina and Monteiro, Mariber and Menick, Jacob and Borgeaud, Sebastian and Brock, Andrew and Nematzadeh, Aida and Sharifzadeh, Sahand and Binkowski, Mikolaj and Barreira, Ricardo and Vinyals, Oriol and Zisserman, Andrew and Simonyan, Karen},
  booktitle={Advances in Neural Information Processing Systems (NeurIPS)},
  year={2022}
}

@article{tishby2000information,
    title={The Information Bottleneck Method},
    author={Tishby, Naftali and Pereira, Fernando C and Bialek, William},
    journal={arXiv preprint physics/0004057},
    year={2000}
}

@inproceedings{katharopoulos2020transformers,
    title     = {Transformers are {RNN}s: Fast Autoregressive Transformers with Linear Attention},
    author    = {Katharopoulos, Angelos and Vyas, Apoorv and Pappas, Nikolaos and Fleuret, Fran\c{c}ois},
    booktitle = {ICML},
    pages     = {5156--5165},
    year      = {2020}
}

@misc{chai2025longerscalinglongsequence,
      title={LONGER: Scaling Up Long Sequence Modeling in Industrial Recommenders}, 
      author={Zheng Chai and Qin Ren and Xijun Xiao and Huizhi Yang and Bo Han and Sijun Zhang and Di Chen and Hui Lu and Wenlin Zhao and Lele Yu and Xionghang Xie and Shiru Ren and Xiang Sun and Yaocheng Tan and Peng Xu and Yuchao Zheng and Di Wu},
      year={2025},
      eprint={2505.04421},
      archivePrefix={arXiv},
      primaryClass={cs.IR},
      url={https://arxiv.org/abs/2505.04421}, 
}
\end{document}